\documentclass[twocolumn]{webofc}

\usepackage[varg]{txfonts}   % Web of Conferences font
\usepackage{hyperref}
\usepackage{url}
\usepackage{cancel}
\usepackage{ulem}
\usepackage{amsmath}
\usepackage{siunitx}

\hypersetup{colorlinks=true,citecolor=blue,urlcolor=blue,linkcolor=blue}
\begin{document}

\title{Modelling the evolution of flow-induced anisotropy of concentrated suspensions}

\author{\firstname{Pappu} \lastname{Acharya}\inst{1}\fnsep\thanks{\email{pappuacharyaphysics@gmail.com}} \and
        \firstname{Romain} \lastname{Mari}\inst{1}\fnsep\thanks{\email{romain.mari@univ-grenoble-alpes.fr}} 
        % etc.
}

\institute{Universit\'e Grenoble Alpes, CNRS, LIPhy, 38000 Grenoble, France}

\abstract{Suspensions, which exhibit complex behaviors such as shear thickening, thinning, and jamming, are prevalent in nature and industry. However, predicting the mechanical properties of concentrated suspensions, in both steady state and the transient regime, remains a significant challenge, impacting product quality and process efficiency. 
In this study, we focus on developing a robust theoretical framework to explain how flow history governs the anisotropy of mechanical responses in suspensions of hard particles under unsteady flow conditions. 
Our starting point is the Gillissen-Wilson constitutive model, which we confront to DEM simulation data of the micro-structure during steady shear, and shear rotations where the shear axis is rotated by a specific angle around the flow gradient direction. We introduce a simple modification to the Gillissen-Wilson model which leads to a model with higher predictive power in steady state and during shear rotations. 
}
\maketitle

\section{Introduction}
\label{intro}

Suspensions in general do not flow like simple fluids. 
Often, their resistance to flow changes in a nonlinear way depending on how fast they are moving. When a suspension becomes thinner and flows more easily as force is applied, it is called shear thinning and conversely, when flow is hampered when force is applied, it is said to be shear thickening. 
Some suspensions however are seemingly simple, with stresses directly proportional to deformation rates, just like in a Newtonian fluid.
This typically happens for the non-Brownian yet Stokesian suspensions that we consider in this paper (and further call simply non-Brownian suspensions), when suspended particles are too large for thermal fluctuations to affect them, but too small for inertia to play a role~\cite{ness_review}.
These suspensions are not Newtonian, however, as for instance they show finite normal stress differences.
This ``partial'' non-Newtonianness gives them a special status of model non-Newtonian material.

Besides normal stress differences, non-Brownian suspensions show flow history dependence, which is best exemplified by a shear reversal experiment~\cite{gadalamaria_1980}: when the direction of flow is suddenly flipped during a simple shear flow, the viscosity instantaneously drops by a proportion of order one, before slowly getting back up to its steady-state value.
This can be more generally seen whenever the direction of shearing is altered~\cite{Blanc2014,Lin2016,Ness2018,Seto2019,blanc2023rheology,acharya2024tacking,acharya2025shear,Agrawal2024,parra2025transient}.
This is because as a suspension is sheared in a particular direction, its microstructure becomes anisotropic, leading to the formation of particle arrangements that result in more contacts or near contacts along the compressional axis~\cite{gadalamaria_1980,Cates98}. 
%\rom{compact along an axis is not optimal wording, maybe smth along the lines of ``more contacts and near contacts along the compressional axis''? should also intreoduce the word anisotropy here} 
This microstructure anisotropy plays a crucial role in determining the overall flow resistance of the suspension. 
When changing shear direction, the initial misalignment between the newly applied shear and the existing particle network leads to a decrease in flow resistance.
The system requires strain to adjust and realigns its microstructure with the new shear direction, restoring the flow resistance to a steady state.
Developing a robust theoretical framework to predict the evolution of the fabric tensor is thus essential for accurately modeling the behavior of suspensions under varying flow conditions.

In this work, we refine the Gillissen-Wilson (G-W) constitutive model \cite{gillissen2018modeling, gillissen2020constitutive} to improve its accuracy in describing the microstructural evolution of suspensions and enhance its predictive performance across different shear flow scenarios. 
%\rom{this is too technical for the intro, we can instead state that while GW is good at capturing the deviatoric part of the steady-state fabric it is not so good with the isotropic component, and here we propose a simple modification to fix this issue.}
While G-W is good at capturing the deviatoric part of the steady-state fabric, we show it is perfectible regarding the isotropic component, and here we propose a simple modification to address this issue.
We demonstrate that this modification also significantly improves the model's predictive capability for the transient microstructure changes occurring when the flow direction is altered. 
We do so using a recently proposed model flow protocol, called shear rotation~\cite{blanc2023rheology, acharya2024tacking, acharya2025shear}.
We validate all our findings through particle-based simulations.

%%%%%%%%%%%%%%%%%%%%%%%%%%%%%%%%%%%%%%%%%%%%%%%%%%%%%%%%%%%%%%%
\begin{figure}[t!]
\centering
\includegraphics[width=1.0\linewidth]{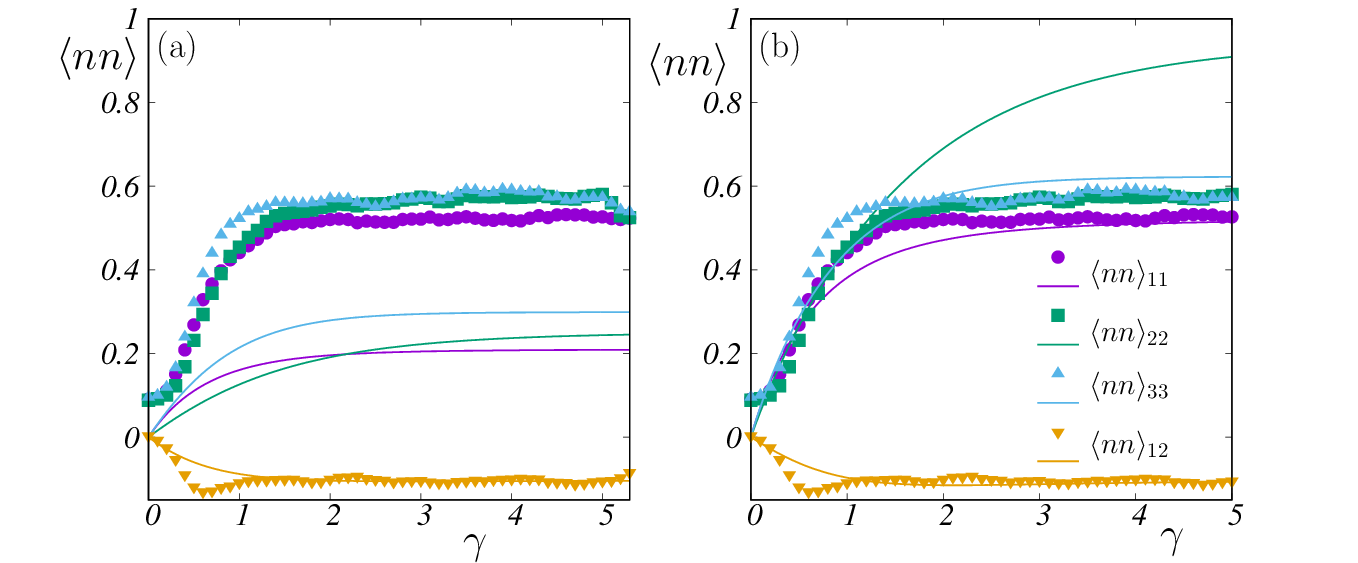}
\caption{Fabric tensor evolution as a function of strain in simulations (symbols) and (a) the standard G-W model Eq.~\ref{eq_gw} (lines) fitted with $\beta \approx 10.9$, (b) modified G-W model Eq.~\ref{eq_Igw2} (lines) fitted with $\beta\approx 9.4$ and $A\approx 42.4$. }
\label{fig_ss}
\end{figure}
%%%%%%%%%%%%%%%%%%%%%%%%%%%%%%%%%%%%%%%%%%%%%%%%%%%%%%%%%%%%%%%

\section{Gillissen-Wilson model}
\label{sec-1}
The Gillissen-Wilson constitutive model describes the strain evolution of the fabric tensor $\langle nn \rangle$.
This tensor is built from the $N_p$ pairs of particles in near contact in the suspension, a near contact being defined as particles whose surfaces are distant by less than a cutoff value $\epsilon$ (here we take $\epsilon$ as $1\%$ of the diameter of the smaller particle in the pair).
Each near contact $\alpha$ is characterized by a center-to-center unit vector $n_\alpha$.
The fabric tensor is then defined as $\langle nn\rangle_{ij}=(1/N)\sum_{\alpha=1}^{N_p} n_{\alpha,i} n_{\alpha,j}$, where $N$ is the number of particles in the suspension and indices $j,k$ represent the components of the tensor. 
We consider the evolution of the fabric tensor under a velocity gradient $L$. 
From $L$, we can define the strain rate tensor $E=(L+L^T)/2$ and the shear rate $\Dot\gamma=\sqrt{2E:E}$.
As the suspension is rate-independent, it is convenient to consider the strain evolution of $\langle nn\rangle$ rather than its time evolution.
In the standard G-W model, the strain evolution
is given by
\begin{multline}
   \partial_\gamma \langle nn \rangle=\hat{L}.\langle nn \rangle + \langle nn \rangle. \hat{L}^T- 2\hat{L}:\langle nnnn \rangle  \\ 
   -\beta \left( \hat{E}_e: \langle nnnn \rangle + \frac{\phi}{15}\left(2\hat{E}_c +Tr(\hat{E}_c)\delta\right)\right),
   \label{eq_gw}   
\end{multline}
with $\hat{L}=L/\Dot{\gamma}$ and $\hat{E}=E/\Dot{\gamma}$, $\delta$ is the identity matrix and $\phi$ is the volume fraction of the suspension.
In the given equation, the terms involving $\hat{L}=L/\Dot{\gamma}$ are usual (upper) convective terms that arise by considering a case where two particles connected by a rigid bond behave like a dumbbell under a velocity gradient $L$.
The strain tensor $\hat{E}$ is divided into two components, $\hat{E}=\hat{E}_e + \hat{E}_c$. $\hat{E}_e$ corresponds to elongation and separates particles, and $\hat{E}_c$, represents compression and brings particles closer. 
This separation is done from the positive and negative eigenvalues ($\lambda$, $-\lambda$ respectively) and normalized eigenvectors ($\hat{v}_{e}$ and $\hat{v}_{c}$ respectively) of $\hat{E}$, which define $\hat{E}_e=\lambda \hat{v}_e\hat{v}_e$ and $\hat{E}_c= -\lambda \hat{v}_c \hat{v}_c$. 
The fourth rank fabric $\langle nnnn \rangle$ is approximated by Hinch and Leal closure~\cite{hinch1976constitutive} and is given by
%\rom{put it also in the main text}
\begin{equation}
\begin{split}
\langle n_i n_j n_k n_l \rangle = 
& -\frac{1}{35} \langle n_m n_m \rangle 
\left( \delta_{ij} \delta_{kl} + \delta_{ik} \delta_{jl} + \delta_{il} \delta_{jk} \right) \\
& + \frac{1}{7} \big( 
\delta_{ij} \langle n_k n_l \rangle + \delta_{ik} \langle n_j n_l \rangle + \delta_{il} \langle n_j n_k \rangle \\
& \quad +  \langle n_i n_j \rangle \delta_{kl}+  \langle n_i n_k \rangle \delta_{jl}+  \langle n_i n_l \rangle \delta_{jk} 
\big).
\end{split}
\end{equation}

Since $\beta$ in Eq.~\ref{eq_gw} influences both the term associated with $\hat{E}_e$ and $\hat{E}_c$, it determines the rate at which contacts form, as well as how the components of the fabric tensor and the stress approach a steady state. 

As we will see shortly, the steady-state solution of the G-W model (obtained as solution of $\partial_\gamma \langle nn \rangle = 0$) underestimates the relative value of the isotropic part of $\langle nn\rangle$ with respect to its deviatoric part.
A potential improvement involves isolating the contribution that acts only on the trace of $\langle nn \rangle$, 
which directly corresponds to the number of contacts per particle. 
The new G-W strain evolution equation will therefore become
\begin{multline}
   \partial_\gamma \langle nn \rangle=\hat{L}.\langle nn \rangle + \langle nn \rangle. \hat{L}^T- 2\hat{L}:\langle nnnn \rangle \\
   -\beta \left( \hat{E}_e: \langle nnnn \rangle + \frac{\phi}{15}2\hat{E}_c\right) - A \frac{\phi}{15}Tr(\hat{E}_c)\delta \, .
   \label{eq_Igw2}
\end{multline}
Setting $A=\beta$, will give the standard G-W model.

With the modified G-W model, the steady-state solution $\langle nn \rangle_{11}^{ss}$ in a simple shear, that is, defining $e_1$, $e_2$ and $e_3$ as, respectively, the flow, gradient and vorticity direction, and setting $L = \dot\gamma e_1 e_2$, is
%\begin{eqnarray}
\begin{align}
\nonumber
\langle nn \rangle_{11}^{ss}&=\frac{\left( 9 \beta \left( 136 - 30 \beta + 9 \beta^2 \right) + 8 A \left( 254 + \beta (-13 + 6 \beta) \right) \right) \phi}{15 \beta \left( 416 + 9 \beta (6 + \beta) \right)} \\[8pt]
\nonumber
\langle nn \rangle_{22}^{ss} &= \frac{\left( A (1864 + 372 \beta + 69 \beta^2) + 6 \beta (-174 + \beta (32 + 3 \beta)) \right) \phi}{15 \beta \left( 416 + 9 \beta (6 + \beta) \right)} \\[10pt]
\nonumber
\langle nn \rangle_{33}^{ss} &= \frac{\left( 464 A + 440 \beta + 92 A \beta + 6 (53 + 8 A) \beta^2 + 81 \beta^3 \right) \phi}{15 \beta \left( 416 + 9 \beta (6 + \beta) \right)} \\[10pt]
\langle nn \rangle_{12}^{ss} &= -\frac{7 \left( A (-8 + \beta) + \beta (-4 + 3 \beta) \right) \phi}{5 \left( 416 + 9 \beta (6 + \beta) \right)}
\end{align}
%\end{eqnarray}
It should be noted that for $A=\beta$, we recover the standard G-W steady-state solution~\cite{gillissen2020constitutive}.
%%%%%%%%%%%%%%%%%%%%%%%%%%%%%%%%%%%%%%%%%%%%%%%%%%%%%%%%%%%%%%%
\begin{figure*}[t!]
\centering
\includegraphics[width=1.0\linewidth]{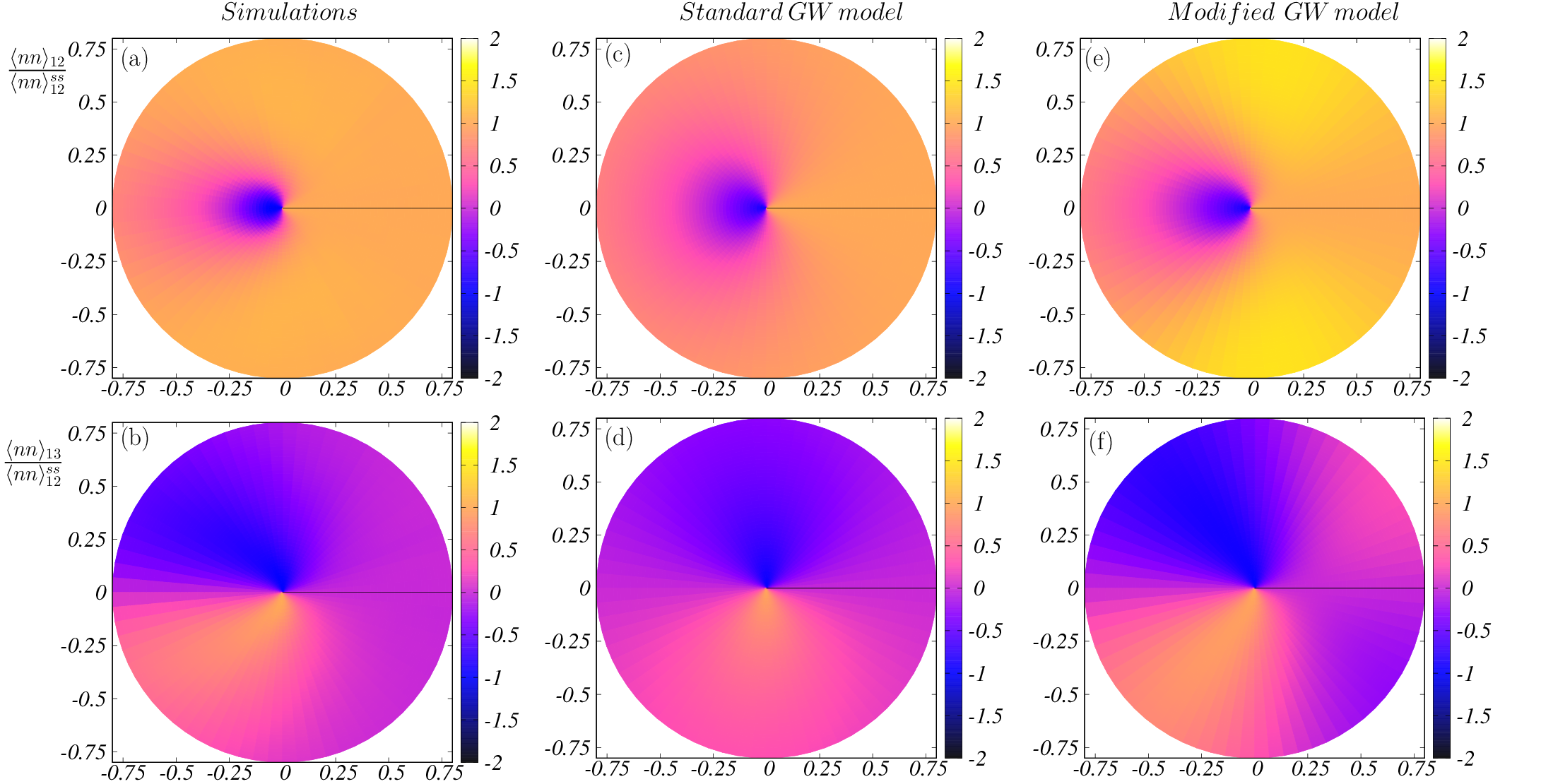}
\caption{Polar plot of the fabric tensor $\langle nn \rangle_{12}/\langle nn \rangle_{12}^{ss}$ and $\langle nn \rangle_{23}/\langle nn \rangle_{12}^{ss}$ with color coding the corresponding values after shear rotation for (a-b) simulations, (c-d) standard G-W model, (e-f) modified G-W model. The radial coordinate is the post-rotation strain and the angular coordinate is the angle of rotation.
%\rom{label $13$ and $23$ to be fixed}
}
\label{fig_color}
\end{figure*}
%%%%%%%%%%%%%%%%%%%%%%%%%%%%%%%%%%%%%%%%%%%%%%%%%%%%%%%%%%%%%%%
\section{Simulation details}
We use a discrete element method~\cite{mari2014shear,trulsson2012transition}, implemented in LAMMPS~\cite{plimpton1995fast,ness2023simulating}.
We simulate a bi-disperse 50:50 mixture of non-Brownian and inertialess stiff spheres with a diameter ratio of 1:1.4 at fixed packing fractions $\phi$, suspended in an iso-dense fluid of viscosity $\eta_f$. 
The system is enclosed within a three-dimensional periodic box, with Lees-Edwards boundary conditions applied to account for shear flow. 
Each particle experiences three primary types of forces and torques: Stokes drag, contact forces including friction, and lubrication forces~\cite{mari2014shear}. 
Here, we use $\phi=0.45$ and friction coefficient $\mu=0.5$. 
We use the strain rate $\dot\gamma$ as the inverse time unit, the small particle diameter $d$ as the length unit and $\rho d^3$ the mass unit, with $\rho$ the particle and fluid density.
With these units, 
we choose particle contact stiffness $k_n=\num{1.25e5}$ so that we remain in the stiff particle limit $k_n \ll P$ with $P$ the measured particle pressure (for the $\phi=0.45$ data presented here, we get $k_n/P \approx \num{e8}$ in steady state). 
We also choose the fluid viscosity $\eta_f=223$ so that the inertial effects remain negligible (Stokes number $\mathrm{St} = \rho d^2 \dot\gamma/\eta_f \ll 1$).

Starting from a random non-overlapping particle configuration, we apply a simple shear up to a total strain of $5$ units. 
We then apply a shear rotation, that is, we rotate the flow $e_1$ and vorticity $e_3$ directions by an angle $\theta$ around the gradient direction $e_2$.

\section{Results}
\subsection{Steady state predictions}

In our simulations, shear is applied along $12$, where 1 is the flow while 2 is the gradient direction. 
As a result, in steady state, by symmetry we have $\langle nn \rangle^{ss}_{13} = \langle nn \rangle^{ss}_{23} = 0$.
% , and only four components \big($\langle nn \rangle^{ss}_{11}$, $\langle nn \rangle^{ss}_{22}$, $\langle nn \rangle^{ss}_{33}$, $\langle nn \rangle^{ss}_{12}$\big) remain nonzero.

In Fig. \ref{fig_ss}, we show the simulation results for the initial startup shear, before shear rotation, and compare them with the predictions of the models.
We pick the $\beta$ parameter in the G-W model, Eq.~\ref{eq_gw}, to fit $\langle nn \rangle_{12}$, as it corresponds to the shearing plane, yielding $\beta \approx 10.9$.
As seen in Fig. \ref{fig_ss}(a), while unsurprisingly the model accurately captures the steady-state value of the fitted component, it does not effectively capture %the transient buildup. Although increasing $\beta$ would speed up the association of particle pairs, it would also come at the expense of compromising the steady-state value. For 
the diagonal components, underestimating their steady-state value.

The modified G-W has two parameters, allowing to fit the steady-state value of $\langle nn \rangle_{12}$ and $\langle nn \rangle_{11}$, yielding $\beta\approx 9.4$ and $A\approx 42.4$. 
The advantage of decoupling the isotropic and deviatoric evolutions is shown in Fig. \ref{fig_ss}(b): the model also captures $\langle nn \rangle_{22}$, only $\langle nn \rangle_{33}$ eludes it as it is now overestimated by the model.

%\section{Steady state and initial transient. Regime in modified G-W model}
%\label{sec-1}

\subsection{Prediction after a shear rotation}
\label{sec-2}
%\rom{use only $1,2,3$.}

The improvement of Eq.~\ref{eq_Igw2} over Eq.~\ref{eq_gw} is illustrated by the behavior of shear stresses under shear rotations.
In Fig.~\ref{fig_color}, we show the simulation results in panels (a)-(b), alongside the G-W predictions in panels (c)-(d) and modified G-W predictions in panels (e)-(f), with the model parameters we picked to fit the steady-state microstructure above. 
In these polar plots, the radial coordinate is the strain $\gamma$ after shear rotation, while the angular coordinate is the angle $\theta$ of shear rotation.
The color encodes the value of $\langle nn \rangle_{12}(\gamma, \theta)/\langle nn \rangle_{12}^{ss}$ [Fig.~\ref{fig_color}(a),(c),(e)] and $\langle nn \rangle_{13}(\gamma, \theta)/\langle nn \rangle_{12}^{ss}$ [Fig.~\ref{fig_color}(b),(d),(f)].
The fact that $\langle nn \rangle_{13}(\gamma, \theta)$ takes finite values during the transients, despite symmetry requiring it vanishes in steady state, is a salient feature of shear rotation which reveals the contribution of the contact stress in the total suspension stress~\cite{blanc2023rheology}.
As expected, $\langle nn \rangle_{12}(\gamma, \theta)$ is even w.r.t. $\theta$ while $\langle nn \rangle_{13}(\gamma, \theta)$ is odd~\cite{blanc2023rheology}.

Comparing Fig.~\ref{fig_color}(c) to Fig.~\ref{fig_color}(a), 
we see that the G-W model overestimates the effect of a shear rotation on the evolution of $\langle nn \rangle_{12}$ for intermediate angles $\pi/4 \lesssim |\theta| \lesssim 3\pi/4$: while simulations show a very quick recovery to the steady-state value for these angles, the G-W model display lengthy transients.
Comparing now Fig.~\ref{fig_color}(d) to Fig.~\ref{fig_color}(b), 
the G-W model now underestimates the effect of a shear rotation on the evolution of $\langle nn \rangle_{23}$ for large angles $\pi/2 \lesssim |\theta| \lesssim \pi$: the recovery predicted by the model is much faster than the one observed in the simulations.

By contrast, the modified model, Eq.~\ref{eq_Igw2}, is significant better at predicting both 
$\langle nn \rangle_{12}(\gamma, \theta)$ and $\langle nn \rangle_{23}(\gamma, \theta)$.
In particular, we see in Fig.~\ref{fig_color}(e) the improvement on $\langle nn \rangle_{12}$ for intermediate angles, and in Fig.~\ref{fig_color}(f) the improvement on $\langle nn \rangle_{23}$ for large angles.
The modified model nonetheless is not devoid of flaws, as for instance it predicts an overshoot on $\langle nn \rangle_{12}$ for angles near $\pi/2$ [Fig.~\ref{fig_color}(e)], which is not present in the simulations.
We are actively working on further refinements to address this limitation.

\section{Conclusion}
\label{sec-1}

We introduced a modification to the G-W model that allows more flexibility as to the relative weights of isotropic versus deviatoric parts of the fabric tensor in steady-state simple shear.
Using particle-based numerical simulations of a dense non-Brownian suspensions, we showed that this refined model indeed shows quantitatively improved predictions in steady state.
%Interestingly, we showed that the refined model is also a clear improvement over the standard G-W model when it comes to the transient dynamics of the fabric tensor under shear rotation, which is a stringent evolution test with a rich phenomenology.
Interestingly, we find that the refined model provides a clear improvement over the standard G-W model in describing the transient dynamics of the fabric tensor under shear rotation. The improved matching in the transient regime constitutes a stringent test of the model’s ability to capture evolution dynamics, as this regime exhibits rich and complex phenomenology.
The obvious next step here is to use the wealth of data from shear rotation simulations to tackle the stress evolution, which within the G-W model is a static relation between the stress tensor $\Sigma$ and the fabric and strain-rate tensors, that is, $\Sigma=\Sigma(\langle nn \rangle, E)$, which is left for future work.
\section*{Acknowledgments}
This project has received funding from the European Union’s Horizon Europe research and innovation programme under the Marie Skłodowska-Curie grant agreement No 101149195.

\let\clearpage\relax
\end{document}